\documentclass[aps,prd,nofootinbib,notitlepage,preprintnumbers,superscriptaddress]{revtex4-1}

\usepackage{amssymb,amsmath,bm,natbib}
\usepackage[dvipsnames]{xcolor}
\usepackage[colorlinks = true,
linkcolor = Maroon,
urlcolor  = Maroon,
citecolor = Maroon]{hyperref}
\usepackage{microtype}
\usepackage{graphicx}
\usepackage{xspace}
\usepackage{bm}
\usepackage{subcaption}
\usepackage{dsfont}

\usepackage{tikz}
\usetikzlibrary{decorations.markings}
\usetikzlibrary{decorations.pathmorphing}
\tikzset{snake it/.style={decorate, decoration=snake}}
\usetikzlibrary{matrix,calc}

\newcommand{\DDs}{\ensuremath{D^0\bar{D}^{*0}}\xspace}
\newcommand{\DDsC}{\ensuremath{\bar D^0{D}^{*0}}\xspace}
\newcommand{\DDpi}{\ensuremath{D^0 \bar D^0 \pi^0}\xspace}
\newcommand{\X}{\ensuremath{X(3872)}\xspace}
\newcommand{\D}{\ensuremath{D^0}\xspace}
\newcommand{\Ds}{\ensuremath{\bar{D}^{*0}}\xspace}

\def\be{\begin{equation}}
	\def\ee{\end{equation}}
\newcommand{\gsim}{\lower.7ex\hbox{$\;\stackrel{\textstyle>}{\sim}\;$}}
\newcommand{\lsim}{\lower.7ex\hbox{$\;\stackrel{\textstyle<}{\sim}\;$}}

\begin{document}
	
	\title{The role of  the pion in  the lineshape of the $X(3872)$}
	
	\author{Angelo~Esposito}
	\email{angelo.esposito@uniroma1.it}
	\affiliation{Dipartimento di Fisica, Sapienza Universit\`a di Roma, Piazzale Aldo Moro 2, I-00185 Rome, Italy}
	\affiliation{INFN Sezione di Roma, Piazzale Aldo Moro 2, I-00185 Rome, Italy}

	\author{Davide~Germani}
	\affiliation{Dipartimento di Fisica, Sapienza Universit\`a di Roma, Piazzale Aldo Moro 2, I-00185 Rome, Italy}
	
	\author{Alfredo~Glioti}
	\affiliation{Universit\'e Paris-Saclay, CNRS, CEA, Institut de Physique Th\'eorique, 91191, Gif-sur-Yvette, France}
	
	\author{Antonio~D.~Polosa}
	\affiliation{Dipartimento di Fisica, Sapienza Universit\`a di Roma, Piazzale Aldo Moro 2, I-00185 Rome, Italy}
	\affiliation{INFN Sezione di Roma, Piazzale Aldo Moro 2, I-00185 Rome, Italy}
	
	\author{Riccardo~Rattazzi}
	\affiliation{Theoretical Particle Physics Laboratory, Institute of Physics, EPFL, 1015 Lausanne, Switzerland}
	
	\author{Michele~Tarquini}
	\affiliation{Walter Burke Institute for Theoretical Physics California Institute of Technology, Pasadena, CA 91125 USA}

	
	\begin{abstract}
		We determine the contribution of long-range pion interactions to  the \X dynamics, assuming it is a loosely bound \DDs molecule. Our result is based on the distorted wave Born approximation  in non-relativistic quantum mechanics. Despite their long-range nature, we find that pion interactions cannot  produce a large and negative effective range. Nonetheless, they introduce imaginary parts. In particular, they contribute to the total decay width of the \X a term associated with, but not precisely corresponding to, the $D^*$ width.
		Our approach can also be applied to the recently discovered $T_{cc}^+$ state.
	\end{abstract}
	
	\maketitle
	
	
	\section{Introduction}
	
	The \X was the first heavy-light  four-quark resonance to be observed. Yet, after almost  two decades, its nature is still up for debate~\cite{Esposito:2016noz,Lebed:2016hpi,Guo:2017jvc,Brambilla:2019esw}. What makes this state particularly interesting, and challenging, is the presence of (at least) two parametric coincidences, or {\it fine tunings}. 
	
	The first tuning is given by the  extreme closeness of the mass of the \X to the \DDs threshold. Indeed, as of today, we only have an upper bound on the distance $B$ from threshold: $B \lesssim 120$~keV~\cite{Braaten:2021iot}. 
	There are two different perspectives on this fact, corresponding to two competing explanations for the nature of the \X.  On the one hand, the \X could be a genuine tetraquark~\cite{Maiani:2004vq,Maiani:2014aja}, i.e. a compact four-quark hadron held together by  gluon mediated forces,  hence  present in the spectrum of QCD at distances of the order of a fermi. In this case, given the very different color configuration of the two-meson state and of the tetraquark, one would  na\"ively expect the natural scale of $B$ to be roughly $\Lambda_{\rm QCD}$. In the face of that expectation, the observed value of $B$ would correspond to a remarkable $1/10^{3}$ fine tuning. However, one observes that all tetraquarks are systematically found within $10-20$ MeV of their corresponding two-meson threshold. There could thus exist a genuine explanation within QCD~\cite{Maiani:2017kyi}, possibly based on the $1/N$ expansion, for this systematic property. That would still single out the \X as a $1/10^2$ outlier in fine tuning space.

	On the other hand, the \X could be a very extended and loosely bound  \DDs state\footnote{As \X is $C$-even, we would actually be dealing with the superposition $\DDs+\DDsC$. Throughout our discussion the proper $C$ quantum number is always understood, even when not properly indicated.}~\cite{Swanson:2003tb,Tornqvist:2004qy,Guo:2017jvc}, a hadronic molecule appearing in the QCD spectrum at distances larger than the fermi, very much like the deuteron. In this case, the closeness to threshold would arise from a tuning of the short distance \DDs interaction, resulting in a large scattering length, $a$ (see, e.g.,~\cite{Braaten:2003he}), and in a correspondingly small binding energy $ B \sim 1/a^2 m_D$. To match the observed value of $B$ one would need $1/a \sim 13$~MeV, corresponding to a mild $1/10$ tuning with respect to the na\"ivest expectation $1/a \sim \Lambda_{\rm QCD}$.

	The second tuning, instead, is the fact that the mass of the \Ds is almost exactly equal to the sum of the masses of the \D and the neutral pion: $m_{D^*} - m_D - m_{\pi} \simeq 7$~MeV~\cite{ParticleDataGroup:2020ssz}. In particular, this implies that, in this channel, the pion mediates long-range interactions, and  that it cannot be integrated out in an effective theory  below the QCD scale.
	
	A promising way of discriminating between the different options for the \X is offered by the study of its lineshape. In particular,  the value of the effective range, $r_0$,  is a good discriminator~\cite{Kang:2016jxw,Esposito:2021vhu,Baru:2021ldu,Kinugawa:2021ykv,Kinugawa:2022ohs,Song:2022yvz,Sazdjian:2022kaf,Mikhasenko:2022rrl,Albaladejo:2022sux,Kinugawa:2023fbf}, as already pointed out by Weinberg in the 60's, when asking the same question for  the deuteron~\cite{Weinberg:1965zz}. On the one hand, for a molecular state, one expects $r_0$ to be controlled by the very size of the bound mesons,  that is $r_0\sim 1 \text{ fm} \sim 1/m_\pi$.\footnote{In our discussion,  we make the rough quantitative identification $1 \text{ fm} \sim 1/m_\pi \sim 1/\Lambda_{\rm QCD}$, even though $m_\pi$ and $\Lambda_{\rm QCD}$ are parametrically distinguished.} This expectation also abides by the modern effective field theory perspective:
	unlike the scattering length $a$, the effective radius $r_0$ is associated with an {\it irrelevant operator} and cannot be tuned larger than the cut-off scale~\cite{Kaplan:1998we}. On the other hand, one finds, by explicit computation, that the presence of
	an interacting compact object close to the two meson threshold produces a negative $r_0$ with an absolute value  larger than $1/m_{\pi}$. Again this is in agreement with the EFT perspective: the only way to enhance an irrelevant operator is to lower the cut-off scale, which in this case is controlled by the separation of the tetraquark from the two meson threshold.
	
	Now, as it was  pointed out in~\cite{Esposito:2021vhu},  a recent LHCb analysis~\cite{LHCb:2020xds}  suggests the second situation for $r_0$ (but see also~\cite{Baru:2021ldu}). However, because of the accidental tuning  we mentioned above, the pion-mediated interaction is also characterized by a large scale. One is thus led to wonder if that could play a role in producing a large and negative effective range.

	In this work, we address this question using a non-relativistic quantum mechanical treatment based on the distorted wave Born approximation. An expression for the effective range for the \X, including the effect of pions, already appeared in~\cite{Braaten:2020nmc}, where a non-relativistic effective theory for the \Ds, the \D and the pion was used. When working at leading order in the small ratio $m_\pi / m_D$, the approach we use here considerably simplifies the problem, bypassing  lengthy calculations. When properly compared, our results are in agreement with Ref.~\cite{Braaten:2020nmc}. 
	
	Comparing to  the most recent experimental data we find that, despite their long-range nature, pion-mediated effects are too weak to generate the large effective radius suggested by observations. However, our result introduces a new structural feature: a complex effective range. Moreover, we find a pion-induced long-range contribution to the  decay width of a molecular \X that does not simply reduce to the decay width of the $D^*\to  D\pi$ process.
	This contribution should be added  to those associated with other (short distance) channels, like for instance $J/\psi\omega$ or $J/\psi\rho$.
	
	We  stress that our analysis can be applied with many similarities to the recently discovered doubly-charmed $T_{cc}^+$ state~\cite{LHCb:2021vvq,LHCb:2021auc}, which is manifestly exotic and shares several features with the \X. In particular, it was pointed out that the effective range of the $T_{cc}^+$ is  negative and much larger than $1/m_\pi$~\cite{Mikhasenko:2022rrl}. This again would appear to speak strongly in favor of a compact tetraquark nature, even though this conclusion was challenged in~\cite{Du:2021zzh}.
	
	
	\section{The effective $DD^*$ Hamiltonian} \label{sec:II}
	
	The LHCb collaboration has recently performed a high statistics analysis of the $B \to K X \to K J/\psi \rho$ process, for events where the invariant mass of the $J/\psi \rho$ pair is close to that of the \X~\cite{LHCb:2020xds}. Plausibly assuming that the $B\to K \DDs$ and the $\DDs \to J/\psi\rho$ vertices originate at  short distance and that they hence are approximately pointlike,   the $B \to K J/\psi \rho$ amplitude results proportional to that for $\DDs \to \DDs$~\cite{Braaten:2005jj}, and can thus be used to extract the \X lineshape.
	
	As already mentioned, since  $m_{D^*}> m_D + m_\pi$, the $D^{*0} \to D^0 \pi^0$ decay is kinematically allowed. However, due to an accidental fine tuning, the available phase space is small  and  all the particles are endowed with  a small velocity. We can therefore describe the relevant dynamics within non-relativistic quantum mechanics, with  two components of the Hilbert space: one describing \DDs (and \DDsC) and the other describing  $\DDpi$. We can represent the Hamiltonian on this two-component Hilbert  space in block diagonal form as
	\begin{align}
		H=\begin{pmatrix} H_{DD^*}& H_I^\dagger \\ H_I& H_{DD\pi}\end{pmatrix} \,,
	\end{align}
	where, in an obvious notation,  $H_{DD^*}$ and $H_{DD\pi}$ describe evolution within the two sectors in the absence of pion interactions, while $H_I$ contains the $D^0 \bar{D}^{*0} \pi$ interaction vertex that couples the two sectors.
	Forward time evolution, and the $S$-matrix, can be  computed using the retarded Green function ${\cal G}_+(E)=(E-H+i\epsilon)^{-1}$.
	As we are only interested in the evolution in the \DDs subspace we only need the Green function reduced to the \DDs block ${\cal G}_+(E)\vert_{\DDs}$. The latter can be conveniently expressed by {\it {integrating out}} the $\DDpi$ component as ${\cal G}_+(E)\vert_{\DDs}=(E-H_{\rm eff}(E)+i\epsilon)^{-1}$
	with the  effective Hamiltonian $H_{\rm eff}(E)$ given by~\cite[e.g.,][]{pethick_smith_2008},
	\begin{align} \label{eq:Heff}
		H_{\rm eff}(E) = H_{DD^*} + H_I^\dagger \frac{1}{E - H_{DD\pi} + i \epsilon} H_I \,.
	\end{align}
	
	The $H_{DD^*}$ Hamiltonian includes the kinetic terms\footnote{As already mentioned, we omit the \DDsC part, which has precisely the same form.} and a pointlike interaction  controlled by a bare coupling, $\lambda_0$,
	\begin{align}
		H_{DD^*} = \frac{{\bm p}_{D^*}^2}{2 m_{D^*}} + \frac{{\bm p}_{D}^2}{2 m_{D}} - \lambda_0 \delta^{(3)}(\bm r) \,,
		\label{hddstar}
	\end{align}
	with $\bm r$ the relative \DDs position.\footnote{To be consistent with power counting, at this order, one should also include higher derivative corrections to the contact \DDs interaction, as explained, for example, in~\cite{Fleming:2007rp,Jansen:2013cba}. Nonetheless, we know that such corrections give a contribution to the effective range of order $m_\pi^{-1}$. Here we are interested in tracking possible contributions substantially larger than that, and we will therefore ignore these corrections. } The  coupling $\lambda_0$ is assumed to bind the \DDs with a large scattering length. Since there is no indication of any such critically large scattering length in the \DDpi system, we model it as non-interacting:
	\begin{align}
		H_{DD\pi} = - \delta + \frac{\bm p_{D,1}^2}{2m_D} + \frac{\bm p_{D,2}^2}{2m_D} + \frac{\bm p_\pi^2}{2m_\pi} \,.
		\label{hddpi}
	\end{align}
	Here $\delta \equiv m_{D^*} - m_D - m_\pi$, and the constant term is due to the fact that, in our notation, all energies are measured with respect to $m_{D^*} + m_D$, and therefore the \DDpi system has a slightly negative mass. 
	
	As far as the interaction Hamiltonian is concerned, instead, it is simpler to write directly its matrix element between states of definite momentum and polarization. At the lowest order in the small pion momentum, the Galilei invariant matrix element can be written as~\cite{Fleming:2007rp,Jansen:2013cba,Braaten:2015tga,Braaten:2020nmc},
	\begin{widetext}
		\begin{align}  \label{eq:matrixelement}
			\langle \bar D(\bm k_1) D(\bm k_2) \pi(\bm q)| H_I | \bar D_\lambda^*(\bm p_1) D(\bm p_2) \rangle = \frac{i g}{2\sqrt{m_\pi}f_\pi m_{D^*}} \left (m_D\bm p_1-m_{D^*} \bm k_1\right ) \! \cdot \! \bm{\varepsilon}_\lambda \, {(2\pi)}^6 \delta^{(3)}(\bm p_1 - \bm k_1 - \bm q) \delta^{(3)}(\bm p_2 - \bm k_2)  \,,
		\end{align}
	\end{widetext}
	where $\bm \varepsilon_\lambda$ is the polarization vector of the \Ds. The same matrix element applies to the charge-conjugated states. It should be stressed that the absence of time-dependent phase factors in Eq.~\eqref{eq:matrixelement} underlies a specific choice of the non-relativistic fields. In the present case that implies the masses appearing in the kinetic terms in Eqs.~\eqref{hddstar}~and~\eqref{hddpi}, but not in the quantity $\delta$, satisfy $m_{D^*}-m_D-m_\pi=0$. Henceforth, we shall  everywhere set $m_{D^*}=m_D+m_\pi$, except when $\delta$ is considered.

	The coupling in Eq.~\eqref{eq:matrixelement} is related to the $D^{*0} \to D^0 \pi^0$ decay width by
	\begin{align}
		\Gamma_* = \frac{g^2 \mu^3}{12\pi f_\pi^2} \,,
		\label{starwidth}
	\end{align}
	where we defined $\mu \equiv \sqrt{2 m_\pi \delta} \simeq 43$~MeV, and $f_\pi \simeq 132$~MeV. Since no experimental value is available for the decay of the neutral $D$-meson, $g$ can be extracted from the width of the charged one,  which gives $g^2\simeq 0.34 $~\cite{Braaten:2015tga}. Notice that this result compares well with  the expectation from {\it {naive dimensional analysis}}~\cite{Manohar:1983md,Georgi:1986kr}, which would roughly give $ g^2 \sim (4\pi f_\pi)^2/N_c m_D^2\sim 0.2$. 
	
\begin{figure}
		\centering
		\resizebox{0.9\textwidth}{!}{
			\begin{tikzpicture}
				\draw[thick, double] (-2,1) -- (-1,1);
				\draw[thick] (-1,1) -- (1,1);
				\draw[thick,double] (1,1) -- (2,1);
				\draw[thick, dashed] (1,1) arc (0: -180: 1);
				\draw[thick] (-2,-1) -- (2,-1);
				
				\node at (-1,1) [circle,fill,inner sep=1.5pt]{};
				\node at (1,1) [circle,fill,inner sep=1.5pt]{};
				
				\node at (-2.5,1) {$\bm x_1, \lambda$};
				\node at (-2.5,-1) {$\bm x_2$};
				\node at (2.5,1) {$\bm y_1, \lambda^\prime$};
				\node at (2.5,-1) {$\bm y_2$};
				
				\node at (-1.7,0.7) {$\bar D^{*0}$};
				\node at (1.7,0.7) {$\bar D^{*0}$};
				\node at (-1.7,-0.7) {$D^{0}$};
				\node at (1.7,-0.7) {$D^{0}$};
				\node at (0,0.7) {$\bar D^0$};
				\node at (0,-0.2) {$\pi^0$};
				
				\draw[thick, double] (5,1) -- (6,1);
				\draw[thick] (6,1) -- (9,1);
				\draw[thick] (5,-1) -- (8,-1);
				\draw[thick,double] (8,-1) -- (9,-1);
				\draw[thick,dashed] (6,1) -- (8,-1);
				
				\node at (6,1) [circle,fill,inner sep=1.5pt]{};
				\node at (8,-1) [circle,fill,inner sep=1.5pt]{};
				
				\node at (5.3,0.7) {$D^{*0}$};
				\node at (8.7,0.7) {$D^{0}$};
				\node at (5.3,-0.7) {$\bar D^{0}$};
				\node at (8.7,-0.7) {$\bar D^{*0}$};
				\node at (7.4,0.1) {$\pi^0$};
				
				\node at (4.5,1) {$\bm x_1, \lambda$};
				\node at (4.5,-1) {$\bm x_2$};
				\node at (9.5,1) {$\bm y_2$};
				\node at (9.5,-1) {$\bm y_1, \lambda^\prime$};
			\end{tikzpicture}
		}
		\caption{Diagrammatic representation of the two transition matrix elements in Eqs.~\eqref{eq:transition}.} \label{fig:diag}
	\end{figure}
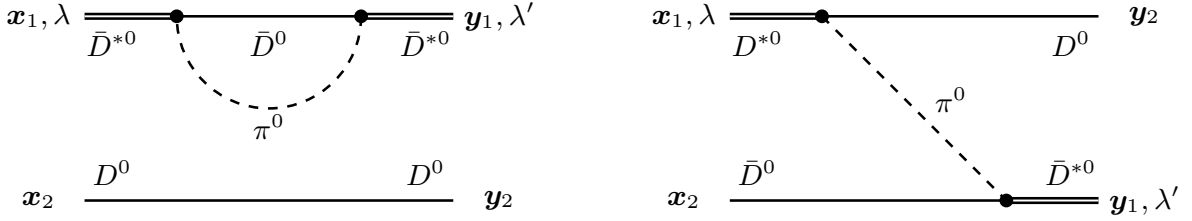

	Now, the term induced by pion exchange (the second) in the effective Hamiltonian in Eq.~\eqref{eq:Heff} consists 
	of two contributions: a $\Ds \D \to \Ds \D$ transition and  a $\Ds \D \to D^{*0} \bar D^0$ one. These correspond to the two diagrams shown in Figure~\ref{fig:diag}, and discussed in~\cite{Jansen:2013cba,Braaten:2015tga,Braaten:2020nmc} in a  quantum field theory setup. A detailed computation of these contributions is reported in Appendix~\ref{app:effpot}. Working in position space, and at leading order in an expansion in $(m_{D^*}-m_D)/m_D\sim m_\pi/m_D \ll 1$, the parts of the two transitions that purely contribute to the $S$-wave read respectively
	\begin{subequations} \label{eq:transition}
		\begin{align}
			\langle \bar D_\lambda^{*}(\bm x_1) D(\bm x_2) | H_I \frac{1}{E - H_{DD\pi} + i \epsilon} H_I | \bar D_{\lambda^\prime}^{*}(\bm y_1) D(\bm y_2) \rangle \simeq {}& \! - i \frac{\Gamma_*}{2} \delta^{(3)}(\bm x_1 - \bm y_1) \delta^{(3)}(\bm x_2 - \bm y_2) \delta_{\lambda \lambda^\prime} \,, \\
			\begin{split}
				\langle D_\lambda^{*}(\bm x_1) \bar D(\bm x_2) | H_I \frac{1}{E - H_{DD\pi} + i \epsilon} H_I | \bar D_{\lambda^\prime}^{*}(\bm y_1) D(\bm y_2) \rangle \simeq {}& \! - \! \left[ \alpha \frac{e^{i\mu r}}{r} + \frac{g^2}{6f_\pi^2} \delta^{(3)}(\bm r) \right] \! \delta^{(3)}(\bm x_1 - \bm y_2) \delta^{(3)}(\bm x_2 - \bm y_1) \delta_{\lambda \lambda^\prime} \,,
			\end{split}
		\end{align}
	\end{subequations}
	where $\bm r = \bm x_1 - \bm x_2$ is the relative $D^*D$ position, and we defined $\alpha \equiv g^2 \mu^2 / (24\pi f_\pi^2) \simeq 5 \times 10^{-4}$. The quantity $\alpha$ measures the weakness of the pion-induced interaction. For instance, Eq.~\eqref{starwidth} reads $\Gamma_*=2\alpha \mu$, indicating that the $D^*$ is actually much narrower than what is implied by the small phase space. The smallness of $\alpha$ simply follows from the derivative nature of pion interactions. Finally, notice that the above leading order matrix elements are independent of $E$: $H_{\rm eff}(E)\equiv H_{\rm eff}$.
	
	Decoupling the motion of the center of mass coordinate, the Schr\"odinger equation for
	the relative \DDs distance $\bm r$ then reads
	\begin{align} \label{eq:Sch}
		H_{\rm eff} \, \psi(r) \equiv 	\left [	-\frac{\nabla^2 }{2\mu_r} - \left(\lambda_0 + \frac{4\pi \alpha}{\mu^2}\right)	\delta^{(3)}(\bm r)  - i \frac{\Gamma_*}{2}  - \alpha \frac{e^{i\mu r}}{r}\right ] \psi(r) = E \, \psi(r) \,,
	\end{align}
	with $\mu_r$ the reduced \DDs mass. The wave function of the \X corresponds to the $C=+1$ combination of the two charge conjugated states, $\psi = \frac{1}{\sqrt{2}} \left( \psi_{\bar D^*\!D} + \psi_{D^* \!\bar D} \right)$. A detailed derivation of the results above can be found again in Appendix~\ref{app:effpot}.
	
	A comment is in order. The effective potential in Eq.~\eqref{eq:Sch} is complex and has an infinite range. This reflects the fact that since the $D^{*0} \to \D \pi^0$ decay is allowed, the intermediate pion in the scattering term can be real, and hence propagate at arbitrarily large distances. Since the reduced \DDs subsystem does not include on-shell pions, the potential is non-Hermitian and unitarity is not manifest. Because of that, as we show in the next sections, there is  an additional correction to the total width of the \X and the effective range turns out to be complex. 
	
	\section{Effects of soft pions on the lineshape of the \X}
	
	We now discuss the consequence of Eq.~\eqref{eq:Sch} on the lineshape of the \X. We perform our study in two steps. First, we consider the scattering problem for positive real energies  close to the (complex) \DDs threshold. Secondly, we study the (unstable) bound state.

	
	\subsection{Scattering states: the effective range} \label{sec:scattering}
	
	The solution of Eq.~\eqref{eq:Sch}, and the corresponding scattering amplitudes, can in principle be found numerically, in complete analogy with~\cite{jackiw1995diverse}, but accounting for the  complex terms in the Hamiltonian.
	However exploiting  the smallness of $\alpha$, one can study the problem analytically working in perturbation theory.
	
	To organize the discussion it is convenient to label the four components of $H_{\rm eff}$ in Eq.~\eqref{eq:Sch} according to $H_{\rm eff}=H_0-i\,\Gamma_*/2+V_s+V_w$, with,
	\begin{align} \label{eq:V}
		H_0=\frac{p^2}{2\mu_r}\,, \qquad V_s(\bm r) = - \left( \lambda_0 +  \frac{4\pi \alpha}{\mu^2} \right)  \delta^{(3)}(\bm{r}) \,,  \qquad V_w(\bm r) = - \, {\alpha} \frac{e^{i \mu r}}{r} \,.
	\end{align}
	As the contribution of pion exchange to $V_s$ is just a redefinition of the bare coupling $\lambda_0$, we can absorb it in $\lambda_0$ and forget about it. The genuine effects of pion exchange are then just the width $\Gamma_*$ and $V_w$. Let us then first consider the limit in which these contributions are neglected, and where the scattering amplitude is purely determined by $H_0+V_s$ alone. Scattering states correspond to the $E>0$ continuum.  Indicating by $k \equiv \sqrt{2\mu_r E}$ the relative \DDs momentum, and using the Lippmann--Schwinger formalism, the amplitude reads
	\begin{align}
		f_s(k)=-\frac{\mu_r}{2\pi}\langle \psi_{0,k}|\left [V_s+V_s\frac{1}{E-H_0-V_s+i\epsilon}V_s\right ]|\psi_{0,k}\rangle=  \frac{1}{-1/a_s - i k} \,,
		\label{LS1}
	\end{align}
	with $\psi_{0,k}(r)=(\sin kr)/kr$ the free $S$-wave solution and with
	$a_s$ the physical scattering length that results from the series of  insertions of $V_s$, after renormalization~\cite{jackiw1995diverse,Braaten:2003he}.\footnote{The Born level term in Eq.~\eqref{LS1} gives $a_s =\mu_r\lambda_0/2\pi$, but every single additional insertion of $V_s$ is linearly UV divergent.} Consider now the effects of $\Gamma_*$ and $V_w$. As $V_w$ is sufficiently localized in space, for sufficiently small $\alpha$, its effect on the amplitude will certainly be treatable as a perturbation. $\Gamma_*$, instead, is not spatially localized: since it does not vanish in the limit of very separated particles, its presence affects the very definition of the asymptotic states and must be thus treated exactly. This is however easily done. By Eq.~\eqref{eq:Sch}, the presence of $\Gamma_*$ simply implies that the asymptotic kinetic energy, $k^2/2\mu_r$, now equals $E+i\Gamma_*/2$, by which the on-shell condition is now that $E+i\Gamma_*/2$ is real and positive. Using  $k \equiv \sqrt{2\mu_r (E+i\Gamma_*/2)}$, Eq.~\eqref{LS1} then simply becomes,
	\begin{align}
		f_s(k)=-\frac{\mu_r}{2\pi}\langle \psi_{0,k}|\left [V_s+V_s\frac{1}{E+i\frac{\Gamma_*}{2}-H_0-V_s+i\epsilon}V_s\right ]|\psi_{0,k}\rangle=  \frac{1}{-1/a_s - i k} \,.
		\label{LS2}
	\end{align}
	In terms of the redefined $k$ the amplitude is the same as before: rather intuitively all that matters is the kinetic energy $k^2/2\mu_r$ of the \DDs system. Notice that the amplitude analytically continued to  $E\to 0$ features  a square root singularity in $\Gamma_*$, consistent with the need for a full resummation of the $\Gamma_*$ insertions~\cite{Jansen:2013cba}.
	
	Consider now  $V_w$. At lowest order in $\alpha$, its effects are captured by the  so-called distorted wave Born approximation~\cite{weinberg2013lectures}, resulting in a correction to the amplitude $f_{DD^*}(k)$ given by (again $k = \sqrt{2\mu_r (E+i\Gamma_*/2)}$ for the rest of this section),
	\begin{align}
		f_{DD^*}(k) = f_s(k) - \frac{\mu_r}{2\pi} \langle \psi_{s,k}^{-} | V_w | \psi_{s,k}^{+} \rangle + \mathcal{O}\left(V_w^2\right) \,,
		\label{distBorn}
	\end{align}
	where $| \psi_{s,k}^{\pm} \rangle$ are the asymptotic states for the Hamiltonian $H_0+V_s-i\Gamma_*/2$~\citep[e.g.,][]{jackiw1995diverse},
	\begin{subequations}
		\begin{align}
			| \psi_{s,k}^{\pm} \rangle={}&\left [1+\frac{1}{\frac{k^2}{2\mu_r} -H_0-V_s\pm i\epsilon}\right ]|\psi_{0,k} \rangle\,,\\
			\psi_{s,k}^{\pm}(r) ={}& \frac{\sin(k r)}{k r} + \frac{1}{-1/a_s \mp i k} \frac{e^{\pm i k r}}{r} = e^{\pm i \delta_s} \frac{\sin(k r+\delta_s)}{k r}\,,
		\end{align}
	\end{subequations}
	where $\delta_s$ is the $S$-wave scattering phase due to $V_s$: $\tan\delta_s = - a_s k$. 
	
	With this at hand, the leading order correction to the scattering amplitude in Eq.~\eqref{distBorn} is controlled by
	\begin{align}
		\langle \psi_{s,k}^{-}| V_w | \psi_{s,k}^{+} \rangle = - \,\alpha \, e^{2i\delta_s}  \int  d^3r \frac{e^{i \mu r}}{r} \frac{\sin^2(kr+\delta_s)}{(kr)^2} \,.
	\end{align}
	The amplitude integral features a  logarithmic UV divergence as a consequence of the $\delta$-function potential that shapes the $\psi^\pm_{s,k}$. Indeed turning off $V_s$, one would have $\delta_s=0$ and no UV divergence. This UV divergence  is associated with short-distance physics which we simply parameterize by introducing  a cutoff at $r=\eta$, where we expect $\eta\sim 1 \text{ fm}$. At leading order as $\eta\mu \to 0$, we then get
	\begin{widetext} 
		\begin{align} \label{eq:epsilon}
			\begin{split}
				\langle \psi_{s,k}^- | V_w | \psi_{s,k}^+ \rangle ={}& -\frac{\pi \alpha}{k^2} \bigg\{ \left( e^{2i \delta_s} -1 \right)^2 \left( \gamma_E -i\frac{\pi}{2}+\log \eta\mu\right ) + \log \left(1- \frac{2k}{\mu} \right) + e^{4i\delta_s} \log\left(1+ \frac{2k}{\mu}\right)  \bigg\} + \mathcal{O}(\eta\mu)\,,
			\end{split}
		\end{align}
	\end{widetext}
	where $\gamma_E$ is the Euler--Mascheroni constant. Notice that, besides the UV divergence controlled by $\eta$, this expression features the standard  IR divergence at $\mu\to 0$ associated with Coulombic interactions, to which our potential reduces in this limit. Moreover, one can check that the result is regular at $k\to 0$, corresponding to the validity of perturbation theory in $\alpha$ even at this point.
	
	For small relative momenta, the inverse amplitude can be expanded as follows,
	\begin{align} \label{eq:finv}
		f_{DD^*}^{-1} = - \frac{1}{a_{R}} - i k + \frac{2 \alpha \mu_r}{\mu^2} \left( \frac{2}{a_R^2 \mu^2} - 1 + \frac{ 8 i }{3a_R \mu} \right) k^2 + \mathcal{O}\big(k^4\big) \,, 
	\end{align}
	where the $\mathcal{O}\big(k^2\big)$ term determines the effective range.  Moreover, for consistency with perturbation theory, we truncated our expression at order $\mathcal{O}(\alpha)$. The UV divergence has been fixed by requiring for the $\mathcal{O}\big(k^0\big)$ to match the physical scattering length, $a_R$. In particular, the relation between $a_s$ and $a_R$ is,
	\begin{align} \label{eq:renorm}
		\frac{a_s}{a_R} = 1 - \frac{2\alpha \mu_r}{\mu} \bigg[ \frac{1}{a_R \mu} + \gamma_E \mu a_R + 2 i + \mu a_R \left( \log(\eta\mu) -i \frac{\pi}{2} \right) \bigg] \,. 
	\end{align}
	
	Since the $\DDs \to \DDs$ scattering cannot be accessed experimentally, the physical scattering length must be obtained from the distance of the \X from the \DDs threshold. At lowest order in $\alpha$ this is $a_R = 1/\sqrt{2 \mu_r B}$ (see Eq.~\eqref{eq:EX} below). Considering the current experimental bound, which for a molecular state implies  $0 \text{ keV} < B \lesssim 100$~keV~\cite{LHCb:2020fvo,LHCb:2020xds}, one finds the following constraints for the real and imaginary parts of the effective range induced by pion exchange:
	\begin{align}
		-0.20 \text{ fm} \lesssim \text{ Re} \,r_0 \lesssim - 0.16 \text{ fm} \,, \qquad 0 \text{ fm} \lesssim \text{ Im} \, r_0 \lesssim 0.17 \text{ fm} \,.
	\end{align}
	As one can see, the pion is too weakly coupled to generate a large and negative effective range in a purely molecular scenario.\footnote{Recall that by  ``large''  one normally means an effective range larger than its natural expected value of roughly $1/m_\pi \simeq 1.4$~fm.} Nonetheless, $r_0$ can now have an imaginary part, even if just a small one.
	
	
	
	\subsection{Bound state: pole of the \X}

	The unperturbed Hamiltonian $H_0+V_s$, besides a continuum of scattering states with positive energy, features,
	for $a_s>0$, one bound state with wave function~\cite{jackiw1995diverse} given by
	\begin{align}
		\psi_X(r) = \frac{1}{\sqrt{2\pi a_s}} \frac{e^{- r / a_s}}{r} \,,
	\end{align}
	and with energy $E_X\equiv -B=- \left(2a_s^2\mu_r \right)^{-1}$. In the molecular hypothesis, this  bound state is the very  \X.
	The pion-mediated interaction now causes a shift in the energy of the molecular \X. The effect of the constant width is readily seen, as before, by moving $\Gamma_*/2$ to the right-hand side in Eq.~\eqref{eq:Sch}, so that $-B=- \left(2a_s^2\mu_r \right)^{-1}$ now equals $E+i\Gamma_*/2$, that is $E_X=- \left(2a_s^2\mu_r \right)^{-1} - i\Gamma_*/2$. The effect of $V_w$ can instead be computed in perturbation theory and, at first order, it gives a shift
	\begin{align}
		\begin{split}
			\Delta E_X ={}& \langle \psi_X | V_\ell | \psi_X \rangle = - \frac{\alpha}{2\pi a_s} \int_{r\geq\eta} d^3r \frac{e^{-\frac{2r}{a_s} + i \mu r}}{r^3}  \,.
		\end{split}
	\end{align}
	The real part of this expression features the same UV divergence encountered in the previous section. Regulating 
	the UV divergence by introducing the same  cut-off length, $\eta$, we can express the final result in terms of the renormalized $a_R$ in Eq.~\eqref{eq:renorm} and of the other physical parameters as
	\begin{align} \label{eq:EX}
		\begin{split}
			E_X = - \frac{1}{2 \mu_r a_R^2} - i \frac{\Gamma_*}{2} - \frac{2\alpha}{a_R^3\mu^2} \bigg[ 1 + 2 i a_R \mu - a_R^2 \mu^2 \log\left( 1 + \frac{2i}{a_R\mu} \right) \bigg]  \,.
		\end{split}
	\end{align}
	As one can see, the term in the square bracket has an imaginary part introducing a correction to the total width of the \X: $\Gamma_X = \Gamma_* + \Delta \Gamma_X$. This  can be interpreted 
	as the effect of binding in  $\Ds \to D\pi$. Indeed  $\Delta \Gamma_X$ vanishes in the unbound case, $a_R\to \infty$.
	Given the experimental constraints on the distance of the \X from threshold, one deduces
	\begin{align}
		0 \text{ keV} \lesssim \Delta \Gamma_X \lesssim 1.9 \text{ keV} \,.
	\end{align}
	Notice that the imaginary part of the term in the square bracket in Eq.~\eqref{eq:EX} vanishes as $1/a_R\mu$ in the limit  $a_R\mu\to \infty$. This further suppresses the maximal value of $\Delta \Gamma_X$, given the allowed minimum of $a_R\mu$ sits around $\sim 3$.
	
	\vspace{1em}
	
	 We conclude by commenting on the comparison between our results and those found in~\cite{Braaten:2020nmc}, where the same problem was studied in an effective field theory. First of all, we notice that our Eq.~\eqref{eq:finv} for the effective range reproduces, at leading order in $m_\pi/m_D \ll 1$, almost exactly the expression for the effective range in~\cite{Braaten:2020nmc}, exception made for the term that does not depend on the scattering length. In~\cite{Braaten:2020nmc} this is written in terms of two finite constants, which the authors call $R_0$ and $F_2$. The former is associated to the UV divergences arising from the short distance contribution to the effective range---i.e. to higher derivative corrections to the contact $\DDs$ interaction---while the latter is associated to an additional UV divergence appearing when including pion effects that are subleading in the $m_\pi / m_D \ll 1$ expansion. As explained in Section~\ref{sec:II},  we neglect the contribution coming from higher dimensional contact interaction and, therefore, to properly compare our results with those of~\cite{Braaten:2020nmc}, we must set $R_0=0$. Moreover, one can show that, at leading order in $m_\pi / m_D \ll 1$, the $F_2$ constant can be calculated. When this is done, the results for both the effective range, Eq.~\eqref{eq:finv}, as well as for the energy of the \X, Eq.~\eqref{eq:EX}, perfectly agree with what found in~\cite{Braaten:2020nmc}. 
Finally, we stress that in~\cite{Braaten:2020nmc} it is also shown that the inclusion of pion-mediated interactions induces a breakdown of the effective range expansion. This breakdown is, however, weak, as it is suppressed by $(m_\pi/m_D)^2$ and, therefore, it is not present in our results.

	
	\section{Conclusions}
	According to basic effective field theory, as synthesized in the so-called  Weinberg criterion~\cite{Weinberg:1965zz},  the effective range of the \X, in the hypothesis of a \DDs molecular state, should be ${\cal O} (1 \text{ fm})$, while, in the hypothesis of a compact hadron,  it could be negative and  with magnitude substantially larger than ${\cal O} (1 \text{ fm})$.  The accidental vicinity of the $\Ds$ to the $D \pi$ threshold, however, introduces a new effective mass scale $\mu\ll 1/{\mathrm{fm}}$ in the dynamics of the system. 
	Even though also the strength of pion interactions  matters, the smallness of $\mu$
	instills the suspicion that $\propto 1/\mu$ contributions to the effective range could affect the above clearcut picture,  also jeopardizing the validity of the results obtained in~\cite{Esposito:2021vhu} and~\cite{Mikhasenko:2022rrl}.  In this study, we showed by explicit computation  that is not the case. Our study  confirms  results previously obtained in the literature~\cite{Jansen:2013cba,Braaten:2020nmc}, but with a technically much simpler approach based on non-relativistic scattering theory  and on the exact solution of three-dimensional $\delta$-function potentials in quantum mechanics.   One essential aspect of our computation is the occurrence of a complex long-range potential from pion exchange, which induces shifts to both real and imaginary parts of all the parameters describing the \X lineshape.
	
	Broadly speaking, the deduction of the effective range of exotic hadrons like the \X or the $T_{cc}^{+}$ from experimental data is a very relevant matter, as it could provide a clear, model-independent determination of their nature. While data are already available, we believe that the current theoretical description of the lineshape of the \X is lacking a systematic treatment, which includes all possible effects (contact interactions, charged thresholds, tetraquark contribution, and so on) in a way that can be readily interpreted from a physical viewpoint. We consider the present study as a first step in that direction, which we plan to keep pursuing in future work.

	
	\begin{acknowledgements}
		R.R. is partially supported by the Swiss National Science Foundation under contract 200020-213104 and through the National Center of Competence in Research SwissMAP. A.E. and A.D.P. thank Hans Werner Hammer for an informative discussion. A.D.P. wishes to thank Adam Szczepaniak and Kevin Ingles for useful clarifications.  We also thank the anonymous referee for providing suggestions on how to compare our results with those of~\cite{Braaten:2020nmc}.
		
	\end{acknowledgements}
	
	
	\appendix
	
	\section{Effective Hamiltonian in position space} \label{app:effpot}
	
	We here show  in more detail how the effective Hamiltonian in Eq.~\eqref{eq:Sch} is obtained. We start by computing the transition matrix elements in Eqs.~\eqref{eq:transition}. Consider  first eq.~(\ref{eq:transition}a)  decomposed in momentum space 
	\begin{align} \label{eq:posspace1}
		\begin{split}
			&\langle \bar D_\lambda^{*}(\bm x_1) D(\bm x_2) | H_I \frac{1}{E - H_{DD\pi} + i \epsilon} H_I | \bar D_{\lambda^\prime}^{*}(\bm y_1) D(\bm y_2) \rangle = \int \frac{d \bm p_1}{(2\pi)^3} \frac{d \bm p_2}{(2\pi)^3} \int \frac{d \bm k_1}{(2\pi)^3} \frac{d \bm k_2}{(2\pi)^3} \\
			& \qquad\qquad\quad  \times e^{i \left( \bm p_1 \cdot \bm y_1 + \bm p_2 \cdot \bm y_2 - \bm k_1 \cdot \bm x_1 - \bm k_2 \cdot \bm x_2 \right)} \langle \bar D_\lambda^{*}(\bm k_1) D(\bm k_2) | H_I \frac{1}{E - H_{DD\pi} + i \epsilon} H_I | \bar D_{\lambda^\prime}^{*}(\bm p_1) D(\bm p_2) \rangle \,.
		\end{split}
	\end{align}
	The momentum space matrix element   can be evaluated using the completeness relations for the $\bar D^0 D^0 \pi^0$ system,
	\begin{align}
		&\langle \bar D_{ \lambda}^{*}(\bm k_1) D(\bm k_2) | H_I \frac{1}{E - H_{DD\pi} + i \epsilon} H_I | \bar D_{\lambda^\prime}^{*}(\bm p_1) D(\bm p_2) \rangle = \int \frac{d\bm q_1}{(2\pi)^3} \frac{d\bm q_2}{(2\pi)^3} \frac{d\bm q_3}{(2\pi)^3} \notag \\
		& \qquad\qquad\qquad  \times \langle \bar D_\lambda^{*}(\bm k_1) D(\bm k_2) | H_I \frac{1}{E - H_{DD\pi} + i \epsilon} | \bar D(\bm q_1) D(\bm q_2) \pi(\bm q_3) \rangle \langle \bar D(\bm q_1)  D(\bm q_2) \pi(\bm q_3) | H_I | \bar D_{\lambda^\prime}^{*}(\bm p_1)D(\bm p_2) \rangle \notag \\
		& \qquad\qquad\quad \simeq  \left( \frac{g}{2\sqrt{m_\pi} f_\pi} \right)^2 \! (2\pi)^3  \delta^{(3)}(\bm k_1 - \bm p_1) \delta^{(3)}(\bm k_2 - \bm p_2) \int d\bm q \frac{\bm q\cdot \bm{\varepsilon}_\lambda \bm q\cdot \bm{\varepsilon}^*_{\lambda^\prime}}{E + \delta - \frac{(\bm p_1 - \bm q)^2}{2m_D} - \frac{\bm p_2^2}{2m_D} - \frac{\bm q^2}{2m_\pi} + i \epsilon } \label{eq:A2} \\
		&\qquad\qquad\quad \simeq \frac{4\pi^3 g^2}{f_\pi^2} \delta^{(3)}(\bm k_1 - \bm p_1) \delta^{(3)}(\bm k_2 - \bm p_2) \int d\bm q \, \frac{\bm q\cdot \bm{\varepsilon}_\lambda \bm q\cdot \bm{\varepsilon}^*_{\lambda^\prime}}{\mu^2 - q^2 + i \epsilon} \,, \notag
	\end{align}
	where we expanded at lowest order in $m_\pi / m_D \ll 1$, keeping in mind the energy $E$ will be of the order of the \DDs kinetic energy, $E \sim p^2 / m_D$. Moreover, we keep $\mu = \sqrt{2 m_\pi \delta}$ fixed, since this is the quantity that critically  controls the coordinate dependence of the pion induce complex potential. Indeed $\mu$ acts as momentum cut off of the effective  \DDs theory with the pion integrated out, and the limit $\mu\to 0$  drastically affects the qualitative behavior of the \DDs system.
	
	The last integral in Eq.~\eqref{eq:A2} is UV divergent, and we choose to regularize it with a cutoff $\Lambda$. Using also the orthonormality of the polarization vectors, $\bm \varepsilon_\lambda \cdot \bm \varepsilon_{\lambda^\prime}^* = \delta_{\lambda \lambda^\prime}$, we obtain,
	\begin{align}
		\begin{split}
			\langle \bar D_{\lambda}^{*}(\bm k_1) D(\bm k_2) | H_I \frac{1}{E - H_{DD\pi} + i \epsilon} H_I | \bar D_{\lambda^\prime}^{*}(\bm p_1) D(\bm p_2) \rangle \simeq{}& - \left[ \frac{16\pi^4 g^2 \Lambda^3}{9 f_\pi^2} + \frac{16\pi^4 g^2 \Lambda \mu^2}{3 f_\pi^2} + (2\pi)^6 i \frac{\Gamma_*}{2} \right] \\
			& \times \delta^{(3)}(\bm k_1 - \bm p_1) \delta^{(3)}(\bm k_2 - \bm p_2) \delta_{\lambda \lambda^\prime} \,.
		\end{split} 
	\end{align}
	The real part is UV divergent and can be absorbed in the physical value of $\delta$ which, with an abuse of notation, we will keep labelling by the same symbol. By plugging the above result  in Eq.~\eqref{eq:posspace1}, we obtain the transition matrix element in position space,
	\begin{align} \label{eq:A4}
		\langle \bar D_\lambda^{*}(\bm x_1) D(\bm x_2) | H_I \frac{1}{E - H_{DD\pi} + i \epsilon} H_I | \bar D_{ \lambda^\prime}^{*}(\bm y_1) D(\bm y_2) \rangle \simeq - i \frac{\Gamma_*}{2} \delta^{(3)}(\bm x_1 - \bm y_1) \delta^{(3)}(\bm x_2 - \bm y_2) \delta_{\lambda \lambda^\prime} \,,
	\end{align}
	where we omitted the UV divergent terms.
	
	With similar manipulations, we can find the matrix element for the transition $\Ds \D \to D^{*0} \bar D^0$ in eq.~(\ref{eq:transition}b) . In position space, it reads (see also~\cite{Suzuki:2005ha}),
	\begin{align}
		&\langle  D_\lambda^{*}(\bm x_1) \bar D(\bm x_2) | H_I \frac{1}{E - H_{DD\pi} + i \epsilon} H_I | \bar D_{\lambda^\prime}^{*}(\bm y_1) D(\bm y_2) \rangle \simeq - \frac{g^2}{2f_\pi^2} \delta^{(3)}(\bm x_1 - \bm y_2) \delta^{(3)}(\bm x_2 - \bm y_1) \! \int \!\frac{d\bm q}{(2\pi)^3} e^{i \bm q \cdot \bm r} \frac{\bm\varepsilon_\lambda \cdot \bm q \bm\varepsilon_{\lambda^\prime}^* \cdot \bm q}{q^2 - \mu^2 - i \epsilon} \notag \\
		& \qquad = \frac{g^2}{8 \pi f_\pi^2} e^{i \mu r} \varepsilon_\lambda^i \varepsilon_{\lambda^\prime}^{*j} \bigg[ \hat r_i \hat r_j \left( \frac{3}{r^3} - \frac{3i\mu}{r^2} - \frac{\mu^2}{r} \right) - \delta_{ij} \left( \frac{1}{r^3} - \frac{i \mu}{r^2} \right) - \frac{4\pi}{3} \delta_{ij} \delta^{(3)}(\bm r)  \bigg] \delta^{(3)}(\bm x_1 - \bm y_2) \delta^{(3)}(\bm x_2 - \bm y_1) \,. \label{eq:exchangeamplitude}
	\end{align}
	At low enough energy, only the $S$-wave channel is relevant. To project the above matrix element to $S$-wave, we average over the $\hat{\bm r}$ direction, effectively setting $\hat r_i \hat r_j \to \frac{1}{3} \delta_{ij}$. This returns,
	\begin{align} \label{eq:A6}
		\begin{split}
			\! \langle  D_\lambda^{*}(\bm x_1) \bar D(\bm x_2) | H_I \frac{1}{E - H_{DD\pi} + i \epsilon} H_I | \bar D_{\lambda^\prime}^{*}(\bm y_1) D(\bm y_2) \rangle \simeq{}& \! - \! \bigg[ \alpha \frac{e^{i \mu r}}{r} + \frac{g^2}{6 f_\pi^2} \delta^{(3)}(\bm r) \bigg] \delta^{(3)}(\bm x_1 - \bm y_2) \delta^{(3)}(\bm x_2 - \bm y_1) \delta_{\lambda \lambda^\prime} \,.
		\end{split}
	\end{align}

	With the transition matrix elements~\eqref{eq:A4} and \eqref{eq:A6} at hand, the Schrodinger equations for the $\bar D^{*0} D^0$ and the $D^{*0} \bar D^0$ become a system of two coupled equations. Decoupling already the center-of-mass motion, they read
	\begin{subequations}
		\begin{align}
			- \bigg[ \frac{\nabla^2}{2\mu_r} + i \frac{\Gamma_*}{2} + \lambda_0 \delta^{(3)}(\bm r) \bigg] \psi_{\bar D^* \! D}(r) - \bigg[ \alpha \frac{e^{i \mu r}}{r} + \frac{g^2}{6 f_\pi^2} \delta^{(3)}(\bm r) \bigg] \psi_{D^* \!\bar D}(r) = {}& E \, \psi_{\bar D^* \! D}(r) \,, \\
			- \bigg[ \alpha \frac{e^{i \mu r}}{r} + \frac{g^2}{6 f_\pi^2} \delta^{(3)}(\bm r) \bigg] \psi_{\bar D^* \! D}(r) - \bigg[ \frac{\nabla^2}{2\mu_r} + i \frac{\Gamma_*}{2} + \lambda_0 \delta^{(3)}(\bm r) \bigg] \psi_{ D^* \! \bar D}(r) ={}& E \, \psi_{D^* \!\bar D}(r) \,.
		\end{align}
	\end{subequations}
	Since the \X is a $C=+1$ state, its wave function as meson molecule is $\psi = \frac{1}{\sqrt{2}} \left( \psi_{\bar D^* \! D} + \psi_{D^* \!\bar D} \right)$. The corresponding Schrodinger equation is obtained by adding together the two above and, indeed, reduces to Eq.~\eqref{eq:Sch}.
	
	Finally, it is interesting to show what  the exchange amplitude computed above look like away from the $m_{D^*} \simeq m_D$ limit. Recalling that, in our notation, $m_{D^*} - m_D = m_\pi$, as well as Eq.~\eqref{eq:matrixelement}, one can perform very similar manipulations (albeit more tedious) and get to,
	\begin{align}
		&\langle  D_\lambda^{*}(\bm x_1) \bar D(\bm x_2) | H_I \frac{1}{E - H_{DD\pi} + i \epsilon} H_I | \bar D_{\lambda^\prime}^{*}(\bm y_1) D(\bm y_2) \rangle = \frac{g^2}{4m_\pi f_\pi^2 m_{D^*}^2} \int \frac{d \bm{p}_1}{(2\pi)^3} \frac{d \bm{p}_2}{(2\pi)^3} \frac{d \bm{k}_1}{(2\pi)^3} \frac{d \bm{k}_2}{(2\pi)^3} \\
		& \qquad\qquad\quad  \times e^{i \left( \bm p_1 \cdot \bm y_1 + \bm p_2 \cdot \bm y_2 - \bm k_1 \cdot \bm x_1 - \bm k_2 \cdot \bm x_2 \right)} (2\pi)^3 \delta^{(3)}(\bm p_1 + \bm p_2 - \bm k_1 - \bm k_2) \frac{\left( m_D \bm k_1 - m_{D^*} \bm p_2 \right) \cdot \bm{\varepsilon}_\lambda^* \left( m_D \bm p_1 - m_{D^*} \bm k_2 \right) \cdot \bm{\varepsilon}_{\lambda^\prime}}{E + \delta - \frac{k_2^2}{2m_D} - \frac{p_2^2}{2m_D} - \frac{(\bm k_1 - \bm p_2)^2}{2m_\pi} + i \epsilon } \,. \notag 
	\end{align}
	When expanding in $m_\pi / m_D \ll 1$, one can easily check that the corrections have the same structure as Eq.~\eqref{eq:exchangeamplitude}, with two gradients, acting either on the potential $e^{i\mu r}/r$ or on the $\delta$-functions, with an overall factor of $m_\pi / m_D$. Except for short distance contributions which we are renormalizing away, both these terms scale exactly as Eq.~\eqref{eq:A6} but with an additional  $m_\pi/ m_D$
	suppression, which makes them negligible.	
	
	\bibliography{biblio}
	
\end{document}